\documentstyle[epsfig,cite,color]{article}
\definecolor{red}{cmyk}{0.0,0.99,1.0,0.0}
\definecolor{green}{cmyk}{0.87,0.25,1.0,0.13}
\definecolor{blue}{cmyk}{0.88,0.77,0.0,0.0}
\definecolor{violet}{cmyk}{0.61,1.0,0.14,0.0}
\newcommand {\Was}{W\c as}
\newcommand {\KKMC}{\hbox{${\cal KK}$}\ MC}
\newcommand {\MISR}[2]{{\cal M}^{\rm ISR (#2)}_{#1}}
\newcommand {\Lf}{{\rm Lf }}
\newcommand {\Sf}{{\rm Sf }}
\newcommand {\Sp}{{\rm Sp }}
\begin{document}
\begin{titlepage}
\begin{flushright}
 {\bf BU-HEPP-04/03 }\\
{\bf April, 2004}\\
\end{flushright}

\begin{center}
{\Large Comparison of Exact Results for the
Virtual Corrections to Bremsstrahlung in $e^+e^-$ Annihilation
at High Energies$^{\dagger}$
}
\end{center}

\vspace{2mm}
\begin{center}
{\bf  S.A. Yost$^{a}$,  C.Glosser$^{b}$, S. Jadach$^{c,d}$ and B.F.L. Ward$^{a}$}\\
\vspace{2mm}
{\em $^a$Department of Physics,\\
  Baylor University of Tennessee, Waco, Texas 76798-7316, USA}\\
{\em $^b$Department of Physics,\\
  Southern Illinois University Edwardsville, Edwardsville, Illinois 62026, USA}\\
{\em $^c$CERN, Theory Division, CH-1211 Geneva 23, Switzerland,}\\
{\em $^d$Institute of Nuclear Physics,
        ul. Radzikowskiego 152, Krak\'ow, Poland,}
\end{center}
\vspace{5mm}
\begin{center}
{Presented by S.A. Yost at LCWS 2004, International Conference on Linear Colliders, Paris, April 19-23, 2004}
\end{center}
\vspace{5mm}
\begin{center}
{\bf   Abstract}
\end{center}
\vspace{10mm}
We have compared the virtual corrections to $e^+ e^- \rightarrow f{\overline f}
+ \gamma$ as calculated by S.\ Jadach, M.\ Melles, B.F.L.\ Ward and S.A.\ Yost
to several other expressions.  The most recent of these comparisons is to
the leptonic tensor calculated by J.H.\ K\"uhn and G.\ Rodrigo for radiative
return.  Agreement is found to within $10^{-5}$ or better, as a fraction of
the Born cross section.  
\vspace{10mm}
\renewcommand{\baselinestretch}{0.1}
\footnoterule
\noindent
{\footnotesize
\begin{itemize}
\item[${\dagger}$]
Work partly supported
by the US Department of Energy Contract  DE-FG05-91ER40627 and by
NATO grant PST.CLG.980342.
\end{itemize}
}
\end{titlepage}
High precision studies of the Standard Model at proposed linear colliders
will require per-mil level control of both the theoretical and experimental
uncertainties in many critical processes to be measured. 
One important contribution 
is the virtual photon correction to the single hard brem\-sstrah\-lung
in $e^+e^-$ annihilations\cite{berends,in:1987,jmwy,hans}. Relevant 
Feynman diagrams are shown in Fig.\ 1.

\begin{figure}[hb]
\begin{center}
\epsfig{file=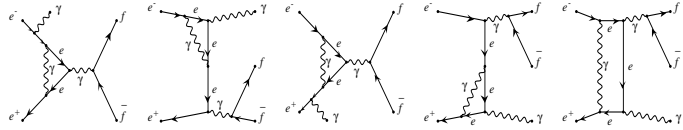,width=90mm}
\end{center}
\vspace{-5mm}
\caption{Feynman diagrams for the virtual ${\cal O}(\alpha^2)$ correction to
the process $e^+e^-\rightarrow f{\overline f}+\gamma$.}
\label{graph1}
\end{figure}

The ${\cal O}(\alpha^2)$ virtual correction to 
single hard brem\-sstrah\-lung can be
expressed in terms of a form factor multiplying the ${\cal O}(\alpha)$
tree level matrix element\cite{jmwy}:
\begin{equation}
\MISR{1}{1} =  {\alpha\over 4\pi} (f_0 + f_1 I_1 + f_2 I_2)\MISR{1}{0}
\end{equation}
where $\MISR{1}{0}$ is the tree-level hard brem\-sstrah\-lung matrix element,
$\MISR{1}{1}$ includes an additional virtual photon, 
and (without mass corrections)
\begin{eqnarray}
f_0 &=& 2\left\{\ln\left({s\over m_e^2}\right) - 1 - i\pi\right\}
+ {r_2\over 1-r_2}
+ {r_2(2+r_1)\over(1-r_1)(1-r_2)} \left\{\ln\left({r_2\over z}\right) 
+ i\pi\right\}
\nonumber\\
&-& \left\{3v + {2r_2 \over 1-r_2}\right\} \Lf_1\left(-v\right)
+ {v \over (1-r_2)}\; R_1(r_1,r_2) + r_2\; R_1(r_2,r_1),
\\
f_1 &=& {r_1 - r_2\over 2(1-r_1)(1-r_2)}
+ {z(1+z)\over 2(1-r_1)^2(1-r_2)}
        \left\{\ln\left(r_2\over z\right) + i\pi\right\}
\nonumber\\
&+& {z\over 1-r_2}\left\{{1\over2} R_1(r_1,r_2) + r_2 R_2(r_1, r_2)\right\}
\nonumber\\
&+& {v\over 4}\left\{R_1(r_1,r_2)\delta_{\sigma,1} + 
R_1(r_2,r_1)\delta_{\sigma,-1}\right\},
\\
f_2 &=& 2 - {1 + z\over 2(1-r_1)(1-r_2)} 
+ {z(r_2 - r_1)\over 2(1-r_1)^2 (1-r_2)}\left\{\ln\left(r_2\over z\right)
        + i\pi\right\}
\nonumber\\
&+& 2 z\; \Lf_2(-v) 
+ {z\over 1-r_2}\left\{{1\over2} R_1(r_1,r_2) + (2-r_2) R_2(r_1, r_2)\right\}
\nonumber\\
&+& {r_1-r_2\over 4}\left\{R_1(r_1,r_2)\delta_{\sigma,1} 
+ R_1(r_2,r_1)\delta_{\sigma,-1}\right\}
\end{eqnarray}
for $\sigma = \lambda_1$, 
with $r_i = 2p_i\cdot k/s$ for momenta $p_1$, $p_2$ of the incoming $e^-$,
$e^+$, $v = r_1 + r_2$ is the fraction of the beam energy radiated into the
hard photon, $z = 1 - v$, and real photon helicity $\sigma$. When $\sigma =
-\lambda_1$, $r_1$ and $r_2$ must be interchanged in eq.\ (2) -- (4). In addition,
we will let $p_3$, $p_4$ label the outgoing $f$, ${\overline f}$ momenta, and
$\lambda_i$ label the helicity of a fermion with momentum $p_i$. The standard
YFS soft virtual photon term $4\pi B_{\rm YFS}$ has been subtracted from
$f_0$.  We make use of functions
\begin{eqnarray}
R_1(x, y) &=& \Lf_1(-x)\left\{\ln\left({1-x\over y^2}\right) - 2\pi i
\right\} 
\nonumber\\
&+& {2(1-x-y)\over 1-x}\; \Sf_1\left({y\over 1-x}, {x(1-x-y)\over 1-x}
\right),
\\
R_2(x, y) &=& 1 - x - y + {1\over 1-x}\left\{\ln\left(y\over 1-x-y\right) + i\pi\right\}
\nonumber\\
&+& \Lf_2(-x)(\ln y + i\pi) 
- (1-x-y)\; \Lf_1(-x)
- {1\over 2}\;\Lf_1^2(-x) 
\nonumber\\
&+& {1-x-y\over (x+y)(1-x)} \left\{x\;\Lf_1\left({-y\over 1-x}\right) 
	- y\;\Lf_2\left({-y\over 1-x}\right)\right\}
\nonumber\\
&+& \left({1-x-y\over 1-x}\right)^2\Sf_2\left({y\over 1-x}, {x(1-x-y)\over 1-x}
\right),
\end{eqnarray}
and $\Lf_n(x)$, $\Sf_n(x, y)$ defined recursively by 
\begin{eqnarray}
&\Lf_0(x) = \ln(1+x),  \quad & 
\Lf_{n+1}(x) = {1\over x}\left(\Lf_n(x) - \Lf_n(0)\right), 
\\
&\Sf_0(x, y) = \Sp(x+y),  \quad & 
\Sf_{n+1}(x, y) = {1\over y}\left(\Sf_n(x,y) - \Sf_n(x,0)\right).
\end{eqnarray}
with $\Sp(x)$ the Spence dilogarithm function. Only the $f_0$ term contributes
to NLL order. The $f_1$ and $f_2$ terms contain spinor coefficients
\begin{eqnarray}
I_1 &=& \sigma \lambda_3 s_{\lambda_1}(p_1,k)
s_{-\lambda_1}(p_2,k)\nonumber\\\nopagebreak
& &\quad\times
{s_{\lambda_3}(p_4,p_2)s_{-\lambda_3}(p_2,p_3)
- s_{\lambda_3}(p_4,p_1)s_{-\lambda_3}(p_1,p_3)
\over s_{-\sigma}(p_1,p_2) s_{-\sigma}(p_3,p_4)
s^2_{\sigma}(p_{21},p_{34})} ,
\\
I_2 &=& \lambda_1\lambda_3{s_{\lambda_1}(p_1,k) s_{-\lambda_1}(p_2,k)
s_{\lambda_3}(p_4,k) s_{-\lambda_3}(p_3,k) \over
s_{-\sigma}(p_1,p_2) s_{-\sigma}(p_3,p_4) 
s^2_{\sigma}(p_{21},p_{34})} ,
\end{eqnarray}
where the spinor product is
$s_{\lambda}(p, q) = {\bar u}_{-\lambda}(p) u_{\lambda}(q)$, and 
$p_{ij} = p_i$ or $p_j$ when $\sigma = \lambda_i$ or $\lambda_j$.
The expressions $f_i$ are equivalent to those in Ref.\ \cite{jmwy}, but with
improved numerical stability in the collinear limits, while the spinor
terms $I_i$ correct misprints in the versions in Ref.\ \cite{jmwy}. Mass 
corrections are added following the method of Ref.\ \cite{bcgktw}, and we
confirmed \cite{jmwy} that all significant mass corrections are included in
this manner.
\begin{figure}[hb]
\begin{center}
\setlength{\unitlength}{0.08mm}
\begin{picture}(1500,1500)
\put(300,250){\begin{picture}(1200,1200)
\put(0,0){\framebox(1200,1200){ }}
\multiput(300,0)(300,0){4}{\line(0,1){25}}
\multiput(0,0)(30,0){41}{\line(0,1){10}}
\multiput(300,1200)(300,0){4}{\line(0,-1){25}}
\multiput(0,1200)(30,0){41}{\line(0,-1){10}}
\put(300,-25){\makebox(0,0)[t]{$0.25$}}
\put(600,-25){\makebox(0,0)[t]{$0.50$}}
\put(900,-25){\makebox(0,0)[t]{$0.75$}}
\put(1200,-25){\makebox(0,0)[t]{$1.00$}}
\multiput(0,133.33)(0,266.67){5}{\line(1,0){25}}
\multiput(0,0.00)(0,26.67){46}{\line(1,0){10}}
\multiput(1200,133.33)(0,266.67){5}{\line(-1,0){25}}
\multiput(1200,0.00)(0,26.67){46}{\line(-1,0){10}}
\put(-25,133){\makebox(0,0)[r]{$0.5\cdot10^{-5}$}}
\put(-25,400){\makebox(0,0)[r]{$1.0\cdot10^{-5}$}}
\put(-25,667){\makebox(0,0)[r]{$1.5\cdot10^{-5}$}}
\put(-25,933){\makebox(0,0)[r]{$2.0\cdot10^{-5}$}}
\put(-25,1200){\makebox(0,0)[r]{$2.5\cdot10^{-5}$}}
\end{picture}}
\put(300,250){\begin{picture}(1200,1200)
\newcommand{\Ra}[2]{\put(#1,#2){\makebox(0,0){\textcolor{red}{$\diamond$}}}}
\newcommand{\E}[3]{\put(#1,#2){\line(0,1){#3}}}
\Ra{15}{62}\E{15}{61}{2}\Ra{45}{74}\E{45}{73}{2}\Ra{75}{80}\E{75}{80}{2}
\Ra{105}{86}\E{105}{85}{2}\Ra{135}{90}\E{135}{89}{2}\Ra{165}{94}\E{165}{93}{2}
\Ra{195}{97}\E{195}{96}{2}\Ra{225}{100}\E{225}{99}{2}\Ra{255}{103}
\E{255}{102}{2}\Ra{285}{106}\E{285}{105}{2}\Ra{315}{108}\E{315}{108}{2}
\Ra{345}{111}\E{345}{110}{2}\Ra{375}{113}\E{375}{112}{2}\Ra{405}{116}
\E{405}{115}{2}\Ra{435}{118}\E{435}{117}{2}\Ra{465}{121}\E{465}{120}{2}
\Ra{495}{123}\E{495}{122}{2}\Ra{525}{126}\E{525}{125}{2}\Ra{555}{129}
\E{555}{128}{2}\Ra{585}{131}\E{585}{130}{2}\Ra{615}{134}\E{615}{133}{2}
\Ra{645}{137}\E{645}{136}{2}\Ra{675}{140}\E{675}{139}{2}\Ra{705}{143}
\E{705}{142}{2}\Ra{735}{146}\E{735}{145}{2}\Ra{765}{150}\E{765}{149}{2}
\Ra{795}{154}\E{795}{153}{2}\Ra{825}{159}\E{825}{158}{2}\Ra{855}{164}
\E{855}{163}{2}\Ra{885}{169}\E{885}{168}{2}\Ra{915}{175}\E{915}{174}{2}
\Ra{945}{183}\E{945}{182}{2}\Ra{975}{192}\E{975}{191}{2}\Ra{1005}{203}
\E{1005}{202}{2}\Ra{1035}{219}\E{1035}{217}{2}\Ra{1065}{241}\E{1065}{239}{2}
\Ra{1095}{275}\E{1095}{274}{2}\Ra{1125}{343}\E{1125}{341}{2}\Ra{1155}{531}
\E{1155}{529}{4}
\end{picture}}
\put(300,250){\begin{picture}(1200,1200)
\newcommand{\Rb}[2]{\put(#1,#2){\makebox(0,0){\textcolor{green}{$\times$}}}}
\newcommand{\E}[3]{\put(#1,#2){\line(0,1){#3}}}
\Rb{15}{60}\E{15}{60}{2}\Rb{45}{71}\E{45}{70}{2}\Rb{75}{76}\E{75}{76}{2}
\Rb{105}{81}\E{105}{80}{2}\Rb{135}{84}\E{135}{83}{2}\Rb{165}{87}\E{165}{86}{2}
\Rb{195}{90}\E{195}{89}{2}\Rb{225}{92}\E{225}{91}{2}\Rb{255}{95}\E{255}{94}{2}
\Rb{285}{97}\E{285}{96}{2}\Rb{315}{99}\E{315}{98}{2}\Rb{345}{101}\E{345}{100}{2}
\Rb{375}{103}\E{375}{102}{2}\Rb{405}{105}\E{405}{104}{2}\Rb{435}{107}
\E{435}{106}{2}\Rb{465}{109}\E{465}{108}{2}\Rb{495}{111}\E{495}{110}{2}
\Rb{525}{113}\E{525}{112}{2}\Rb{555}{115}\E{555}{114}{2}\Rb{585}{118}
\E{585}{117}{2}\Rb{615}{120}\E{615}{119}{2}\Rb{645}{123}\E{645}{122}{2}
\Rb{675}{125}\E{675}{125}{2}\Rb{705}{128}\E{705}{128}{2}\Rb{735}{132}
\E{735}{131}{2}\Rb{765}{135}\E{765}{134}{2}\Rb{795}{140}\E{795}{139}{2}
\Rb{825}{144}\E{825}{143}{2}\Rb{855}{149}\E{855}{148}{2}\Rb{885}{155}
\E{885}{154}{2}\Rb{915}{162}\E{915}{161}{2}\Rb{945}{171}\E{945}{170}{2}
\Rb{975}{181}\E{975}{180}{2}\Rb{1005}{194}\E{1005}{193}{2}\Rb{1035}{212}
\E{1035}{211}{2}\Rb{1065}{239}\E{1065}{238}{2}\Rb{1095}{281}\E{1095}{280}{2}
\Rb{1125}{363}\E{1125}{361}{4}\Rb{1155}{588}\E{1155}{586}{4}
\end{picture}}
\put(300,250){\begin{picture}(1200,1200)
\newcommand{\Rc}[2]{\put(#1,#2){\makebox(0,0){\textcolor{blue}{$\star$}}}}
\newcommand{\E}[3]{\put(#1,#2){\line(0,1){#3}}}
\Rc{15}{70}\E{15}{69}{2}\Rc{45}{85}\E{45}{84}{2}\Rc{75}{95}\E{75}{93}{4}
\Rc{105}{103}\E{105}{101}{4}\Rc{135}{111}\E{135}{108}{4}\Rc{165}{118}
\E{165}{115}{6}\Rc{195}{121}\E{195}{117}{6}\Rc{225}{127}\E{225}{123}{8}
\Rc{255}{134}\E{255}{130}{8}\Rc{285}{141}\E{285}{136}{10}\Rc{315}{147}
\E{315}{142}{10}\Rc{345}{152}\E{345}{147}{12}\Rc{375}{156}\E{375}{150}{12}
\Rc{405}{159}\E{405}{152}{14}\Rc{435}{163}\E{435}{156}{14}\Rc{465}{174}
\E{465}{166}{16}\Rc{495}{180}\E{495}{171}{16}\Rc{525}{183}\E{525}{174}{18}
\Rc{555}{184}\E{555}{175}{20}\Rc{585}{189}\E{585}{179}{20}\Rc{615}{193}
\E{615}{182}{22}\Rc{645}{198}\E{645}{187}{22}\Rc{675}{207}\E{675}{195}{24}
\Rc{705}{220}\E{705}{207}{26}\Rc{735}{227}\E{735}{214}{26}\Rc{765}{235}
\E{765}{221}{28}\Rc{795}{240}\E{795}{225}{30}\Rc{825}{260}\E{825}{244}{32}
\Rc{855}{277}\E{855}{260}{34}\Rc{885}{288}\E{885}{270}{36}\Rc{915}{311}
\E{915}{292}{38}\Rc{945}{338}\E{945}{317}{40}\Rc{975}{380}\E{975}{358}{44}
\Rc{1005}{426}\E{1005}{403}{48}\Rc{1035}{495}\E{1035}{469}{52}\Rc{1065}{562}
\E{1065}{534}{56}\Rc{1095}{669}\E{1095}{637}{62}\Rc{1125}{890}\E{1125}{855}{72}
\end{picture}}
\put(300,250){\begin{picture}(1200,1200)
\newcommand{\Rd}[2]{\put(#1,#2){{\textcolor{violet}{\bf\circle{20}}}}}
\newcommand{\E}[3]{\put(#1,#2){\line(0,1){#3}}}
\Rd{15}{77}\E{15}{76}{2}\Rd{45}{97}\E{45}{96}{2}\Rd{75}{111}\E{75}{110}{2}
\Rd{105}{123}\E{105}{122}{2}\Rd{135}{134}\E{135}{133}{2}\Rd{165}{143}
\E{165}{142}{2}\Rd{195}{152}\E{195}{151}{2}\Rd{225}{160}\E{225}{158}{2}
\Rd{255}{167}\E{255}{166}{2}\Rd{285}{175}\E{285}{174}{2}\Rd{315}{182}
\E{315}{180}{2}\Rd{345}{188}\E{345}{187}{2}\Rd{375}{195}\E{375}{194}{2}
\Rd{405}{202}\E{405}{201}{2}\Rd{435}{208}\E{435}{207}{2}\Rd{465}{215}
\E{465}{213}{2}\Rd{495}{221}\E{495}{220}{2}\Rd{525}{227}\E{525}{226}{2}
\Rd{555}{234}\E{555}{233}{2}\Rd{585}{241}\E{585}{239}{2}\Rd{615}{248}
\E{615}{246}{2}\Rd{645}{254}\E{645}{253}{2}\Rd{675}{261}\E{675}{260}{2}
\Rd{705}{268}\E{705}{266}{2}\Rd{735}{275}\E{735}{274}{2}\Rd{765}{283}
\E{765}{281}{2}\Rd{795}{291}\E{795}{290}{2}\Rd{825}{300}\E{825}{299}{2}
\Rd{855}{309}\E{855}{308}{2}\Rd{885}{320}\E{885}{318}{2}\Rd{915}{331}
\E{915}{329}{2}\Rd{945}{344}\E{945}{342}{2}\Rd{975}{359}\E{975}{357}{2}
\Rd{1005}{376}\E{1005}{375}{4}\Rd{1035}{399}\E{1035}{397}{4}\Rd{1065}{431}
\E{1065}{429}{4}\Rd{1095}{479}\E{1095}{477}{4}\Rd{1125}{567}\E{1125}{565}{4}
\Rd{1155}{801}\E{1155}{799}{6}
\end{picture}} 
\put(0,80){Figure 2: Comparisons of NNLL results in a \KKMC\ run of $10^8$ events.}
\put(300,250){\begin{picture}(1200,1200)
\put(500,1050){\makebox(0,0)[r]{\Large
${\bar\beta_1}^{(2)}-{\bar\beta_{1\rm NLL}}^{(2)}
\over{\bar\beta_{\rm Born}}$}}
\put(600,40){\makebox(0,0)[b]{\Large $v_{\max}$}}
\put(200,800){\begin{picture}(300,400)
\put(0,0){\makebox(0,0)[l]{\textcolor{red}{$\diamond$}}}
\put(90,0){\makebox(0,0)[l]{JMWY}}
\put(0,-120){\makebox(0,0)[l]{\hspace{-1pt}\textcolor{green}{$\times$}}}
\put(90,-120){\makebox(0,0)[l]{IN}}
\put(0,-240){\makebox(0,0)[l]{\textcolor{blue}{$\star$}}}
\put(90,-240){\makebox(0,0)[l]{BVNB}}
\put(0,-360){\makebox(0,0)[l]{\textcolor{violet}{\hspace{2pt}\circle{15}}}}
\put(90,-360){\makebox(0,0)[l]{KR}}
\end{picture}}\end{picture}}
\end{picture}
\end{center}
\end{figure}

Fig.\ 2 shows a comparison of four expressions for the sub-NLL 
virtual photon contribution to the $\bar\beta_1^{(2)}$ distribution at a CMS energy
of 200 GeV, 
with $f{\overline f} = \mu^-\mu^+$.  The NLL contribution calculated in 
Ref.\ \cite{jmwy} has been subtracted in each case.  The figure compares
our exact result JMWY  
in Ref.\ \cite{jmwy}, the result IN of Ref.\ \cite{in:1987}, 
the result BVNB of Ref.\ \cite{berends}, 
the new result KR of Ref.\ \cite{hans}. 
The first two comparisons were 
included in Ref.\ \cite{jmwy}, where good agreement was found. In fact, both
expressions were shown to be analytically identical to ours at NLL order.
However,  neither of the comparisons in Refs.\ \cite{berends,in:1987}
is fully differential with mass corrections.
The result of Ref.\ \cite{hans} is the only comparison which is fully
differential and includes mass corrections, allowing a complete test of 
the sub-NLL terms in eq.\ (1).

All of the results agree to within
$0.4\times 10^{-5}$ for cuts below $v_{\rm max} = 0.75$.  
For cuts between 0.75 and .95,
the results agree to within $0.5\times 10^{-5}$, except for the result
of Ref.\ \cite{berends}.  
These results are consistent with a total precision tag
of $1.5\times 10^{-5}$ for our ${\cal O}(\alpha^2)$
correction $\bar\beta_1^{(2)}$ for an energy cut below $v_{\rm max}=0.95$. The NLL
effect, which has been implemented in the \KKMC\ \cite{kkmc:2001}, is adequate 
alone to within $1.5\times 10^{-5}$ for cuts below 0.95.
More details on the comparisons can be found in Ref.\ \cite{jgwy}.

These comparisons show that
we now have a firm handle on the precision tag
for an important part of the complete ${\cal O}(\alpha^2)$ corrections
to the $f{\overline f}$ production process needed for precision
studies of such processes in the final LEP2 data analysis, in the
radiative return at $\Phi$ and B-Factories, and in the
future TESLA/LC physics. 

\label{references}

\end{document}